\documentclass[aps,twocolumn,showpacs,preprintnumbers,superscriptaddress,amsmath,amssymb]{revtex4}

\usepackage[dvips]{graphicx} 

\usepackage[english]{babel}
\usepackage{blindtext}
\usepackage{dcolumn}
\usepackage{bm}
\usepackage{ulem}
\usepackage[dvips]{color} 

\newcommand{\ie}{{\it i.e. }}
\newcommand{\eg}{{\it e.g. }}
\newcommand{\cf}{{\it cf. }}

\newcommand{\co}[2]{\ifcase #1 \or #2 \fi}

\newcommand{\bscco}{Bi$_{2}$Sr$_{2}$CaCu$_{2}$O$_{8}$\,}

\newcommand{\micron}{$\,\mu$m}
\newcommand{\celsius}{\,$^\circ$C}
\newcommand{\angstrom}{\,$\mathring{A}$}

\newif\ifnote

\notetrue                   %

\begin{document}

\DeclareGraphicsExtensions{.eps}


\title{Hot spots and waves in Bi$_{2}$Sr$_{2}$CaCu$_{2}$O$_{8}$\, intrinsic Josephson junction stacks - a study by Low Temperature Scanning Laser Microscopy }

\author{H.B.~Wang}
\affiliation{National Institute for Materials Science, Tsukuba 3050047, Japan}
\author{S.~Gu\'{e}non}
\affiliation{
Physikalisches Institut --
Experimentalphysik II
and
Center for Collective Quantum Phenomena,
Universit\"{a}t T\"{u}bingen,
Auf der Morgenstelle 14,
D-72076 T\"{u}bingen,
Germany
}
\author{J.~Yuan}
\affiliation{National Institute for Materials Science, Tsukuba 3050047, Japan}
\author{A.~Iishi}
\affiliation{National Institute for Materials Science, Tsukuba 3050047, Japan}
\author{S.~Arisawa}
\affiliation{National Institute for Materials Science, Tsukuba 3050047, Japan}
\author{T.~Hatano}
\affiliation{National Institute for Materials Science, Tsukuba 3050047, Japan}
\author{T.~Yamashita}
\affiliation{National Institute for Materials Science, Tsukuba 3050047, Japan}
\author{D.~Koelle}
\affiliation{ Physikalisches Institut -- Experimentalphysik II and Center for Collective
Quantum Phenomena, Universit\"{a}t T\"{u}bingen, Auf der Morgenstelle 14, D-72076
T\"{u}bingen, Germany }
\author{R.~Kleiner}
\affiliation{
Physikalisches Institut --
Experimentalphysik II
and
Center for Collective Quantum Phenomena,
Universit\"{a}t T\"{u}bingen,
Auf der Morgenstelle 14,
D-72076 T\"{u}bingen,
Germany
}

\date{\today}

\begin{abstract}
Recently, it has been shown that large stacks of intrinsic Josephson  junctions in
Bi$_{2}$Sr$_{2}$CaCu$_{2}$O$_{8}$\, emit synchronous THz radiation, the synchronization presumably triggered by a
cavity resonance. To investigate this effect we use Low Temperature Scanning Laser
Microscopy to image electric field distributions. Apart from verifying the appearance of
cavity modes at low bias we find that, in a high input power regime, standing-wave
patterns are created through interactions with a hot spot, possibly pointing to a new
mode of generating synchronized radiation in intrinsic Josephson junction stacks.
\end{abstract}


\maketitle
Terahertz (THz) physics and technology still lacks good active devices (\eg for nondestructive materials diagnostics or chemical and biological sensing), a problem often referred to as the "Terahertz gap" \cite{Ferguson02}.
Some high temperature superconductors such as \bscco (BSCCO) intrinsically form stacks of Josephson junctions that may operate in this regime.
Recently, synchronous THz emission of almost 1000 of such junctions has been demonstrated \cite{Ozyuzer07}.

When a dc voltage $V$ is applied to a Josephson junction the supercurrent oscillates at a frequency $f=V/\Phi_{0}$, where $\Phi_{0}$ is the flux quantum ($\Phi_{0}^{-1} \approx 483.6$\,GHz/mV).
Unfortunately, $f$ is limited by the superconducting energy gap, restricting operation of conventional Josephson junctions to frequencies below 700\,GHz or so.
Further, a single Josephson junction produces a small output power in the nanowatt or even picowatt range.
Cuprate superconductors, having higher energy gaps, at least in principle allow to operate Josephson devices up to the THz regime, although in reality many junction types have a strong internal damping, again restricting their usability to the sub-THz regime.
The exception are intrinsic Josephson junctions that are naturally formed by the crystal structure \cite{Kleiner92, Yurgens00}.

High frequency emission of unsynchronized intrinsic junctions has been observed up to 0.5 THz \cite{Batov06}.
Various strategies were studied to achieve synchronized THz radiation from intrinsic junction stacks.
These studies included the use of shunting elements in parallel to small sized stacks \cite{Wang00, Grib06,
Madsen04}, the excitation of Josephson plasma oscillations via heavy quasiparticle injection \cite{Lee00,
Kume99} or the investigation of stimulated emission due to quantum cascade processes \cite{Krasnov06}.
The strategy perhaps investigated most intensively considered the generation of collective Josephson oscillations by a lattice of moving Josephson vortices, exciting electromagnetic cavity resonances inside the stack, see e.g. \cite{Kleiner94, Kleiner00, Ustinov98, Machida00, Heim02, Fujino02, Bae07, Bae06, Clauss04, Wang06}.
Here, typically stacks or arrays of stacks with lateral dimensions of a few microns or smaller, consisting of
some 10 junctions have been studied, perhaps with modest success.
By contrast, in the recent experiment \cite{Ozyuzer07} much larger structures consisting of almost 1000 junctions having lateral dimensions in the 100\micron\ range have been used to observe coherent off-chip THz radiation with an output power of some  $\mu$W.
Phase synchronization was presumably achieved via the excitation of cavity resonances.

This experiment immediately triggered activities to understand the mechanism of THz generation in large
intrinsic junction stacks \cite{Bulaevskii07, Koshelev08, Koshelev08b, Lin08}.
%
\begin{figure}[tb]
\begin{center}
\includegraphics[width=0.9\columnwidth,clip]{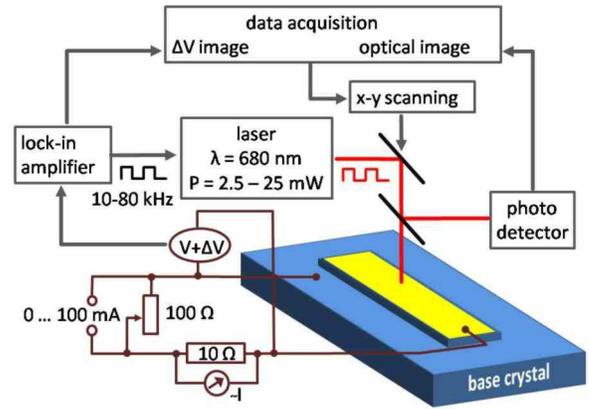}
\end{center}
\caption{(color online). Principle of Low Temperature Scanning Laser Microscopy (LTSLM). The beam of a diode laser ($\lambda=680$ nm) is deflected by a scanning unit and focused onto the surface of the sample. In order to provide a load line for stable operation, the mesa (yellow structure) is biased using a current source and variable resistor in parallel to the mesa.  The local heating by 0.3-3 K due to the laser causes a response $\Delta$V serving as the contrast for the LTSLM image. In order to improve the signal-to-noise ratio the amplitude of the laser beam is modulated at a frequency 10-80\,kHz and analyzed by a lock-in amplifier.
\label{fig:1}}
\end{figure}
Here we report on a study of such structures using Low Temperature Scanning Laser Microscopy (LTSLM), \cf Fig.\ref{fig:1} and Ref. \cite{Peschka02}.
The method allows to visualize electric field and current distributions with a lateral resolution of about 1.5 \micron.
Standing electromagnetic waves have been imaged this way.
Furthermore, the large structures used in \cite{Ozyuzer07} - even larger structures have been suggested \cite{Bulaevskii07} - are prone to substantial heating.
LTSLM is also able to shine light on such heating effects.

For the experiments \bscco (BSCCO) single crystals were grown using the floating zone technique in a four lamp arc-imaging furnace, and annealed at 600\celsius\ and 1 atm (argon 99\% and oxygen 1\%).
A $T_{c}$ of about 83 K is typical for our samples, which are slightly under-doped.
A gold layer 600\angstrom\ thick was sputtered onto the surface of a cleaved BSCCO single crystal.
Noticeably, the single crystals were cleaved in vacuum to have a good contact between the gold and the single crystal.
Then conventional photolithography was used to define the sizes of a mesa structure in
the $a$-$b$ plane.
The length of the mesas was 330\micron, the width varied between 30\micron\ and 80\micron.
By using Ar ion milling, the samples were etched down to yield a mesa thickness of
1\micron\ along the $c$-axis, implying that there are about 670 intrinsic junctions
involved in the stack.
Polyimide was used to surround the edge of the mesa as an insulator, and a gold wire was attached to the mesa by silver paste.
Other gold wires were connected to the big single crystal pedestal as grounds.
In order to provide a load line for stable operation, the mesas were biased using a current source and variable resistor in parallel to the mesa, \cf Fig.\ref{fig:1}.  In total we measured 6 mesas on 3 different crystals.
%

\begin{figure*}
  \includegraphics[width=1.4\columnwidth,clip]{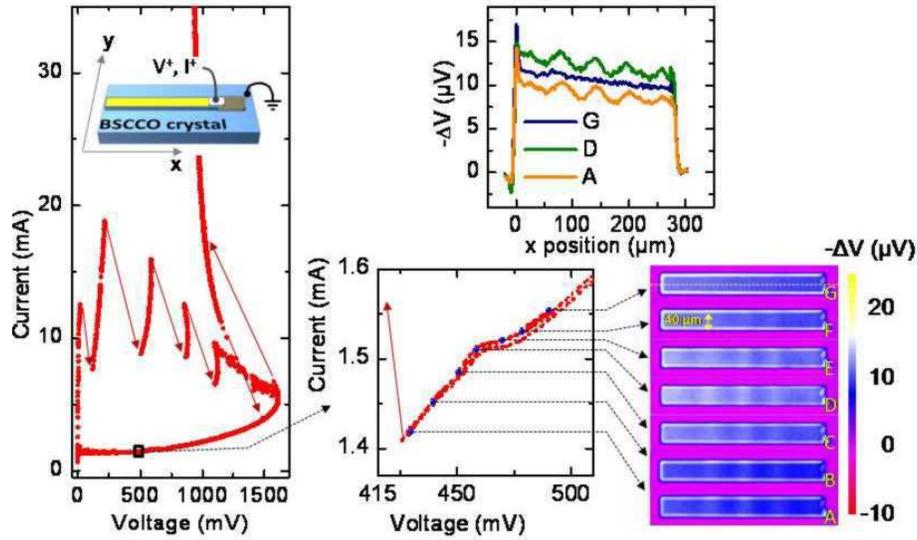}
  \caption{(double column): (color online) Current voltage characteristic (IVC) of a 40 x 330\,$\mu m^{2}$ large BSCCO mesa, measured at 25\,K, together with LTSLM data. Left graph displays the IVC on large current and voltage scales. The inset shows the geometry of the device. The lower graph in the middle shows an enlargement of the IVC in the region where LTSLM images (A) to (G) were recorded. Solid arrows denote jumps in the IVC. The upper graph on the right shows for selected bias currents, corresponding to images (A), (D) and (G), line scans along the long side of the mesa at half width, \cf dashed line in image (G). }
\label{fig:2}
\end{figure*}
We first discuss data of a 40\micron\ wide mesa. Figure \ref{fig:2} shows the current voltage characteristic (IVC) of this sample at 25\,K together with LTSLM data, taken in a bias regime where THz emission was detected in \cite{Ozyuzer07}.
For voltages above 480\,mV the LTSLM images are smooth, \cf image (G), except for an enhanced response  $\Delta V$ along the edge of the mesa.
Here, the laser warms up the mesa more effectively than in the inner region where it is attenuated by the gold layer covering the mesa surface.
The negative sign of  $\Delta V$ is consistent with the fact that the (subgap) resistance of the junctions decreases with increasing temperature.
At lower voltages, \cf images (A) to (F), a wave-like structure appears which is very reminiscent of images of standing electromagnetic waves (cavity resonances), as made by Low Temperature Scanning Electron Beam Microscopy (LTSEM) both for conventional Josephson junctions \cite{Mayer91} and intrinsic Josephson junction stacks \cite{Clauss04}.
Here, in essence, the beam induced temperature rise lowers the quality factor of the resonance, in turn changing the voltage across the junctions.
The effect is strongest when the beam heats up the antinodes of the standing wave and weakest when the beam position is at its nodes (the same physics occurs in LTSLM).
Thus the period in the LTSLM image equals half the wave length $\lambda$ ($\approx$120\micron\ in our case) of the cavity mode.
The images were taken in a state where 60-70\,\% of all junctions were resistive while the others carried no dc voltage, as estimated from the ratio of the voltages on the branch investigated and the outermost branch.
The average voltage per junction, say for image (D), is about 1 mV; using Josephson's relation we estimate $f\approx0.5-0.6$ THz and find a mode velocity $c=f\cdot\lambda\approx(6-7)\cdot10^{7}$\,m/s, which is close to the highest possible value $c_{0}/n\approx 9\cdot10^{7}$\,m/s, where $c_{0}$ is the vacuum speed of light and $n\approx3.5$ is the far infrared diffraction index \cite{Ozyuzer07}.
Thus, the pattern imaged is consistent with the THz emission observed in \cite{Ozyuzer07}, although in our case we see that the wave forms along the long side (in \cite{Ozyuzer07} the fundamental cavity mode with one half-wave along the short side seemed to have formed).

For this mesa we have seen similar patterns on several branches on the IVC (\ie with different numbers of junctions in the resistive state).
However, for the other mesas we could not find corresponding signals.
This kind of cavity modes seems to be difficult to excite.
We next focus on higher bias currents.
\begin{figure*}
  \includegraphics[width=1.4\columnwidth,clip]{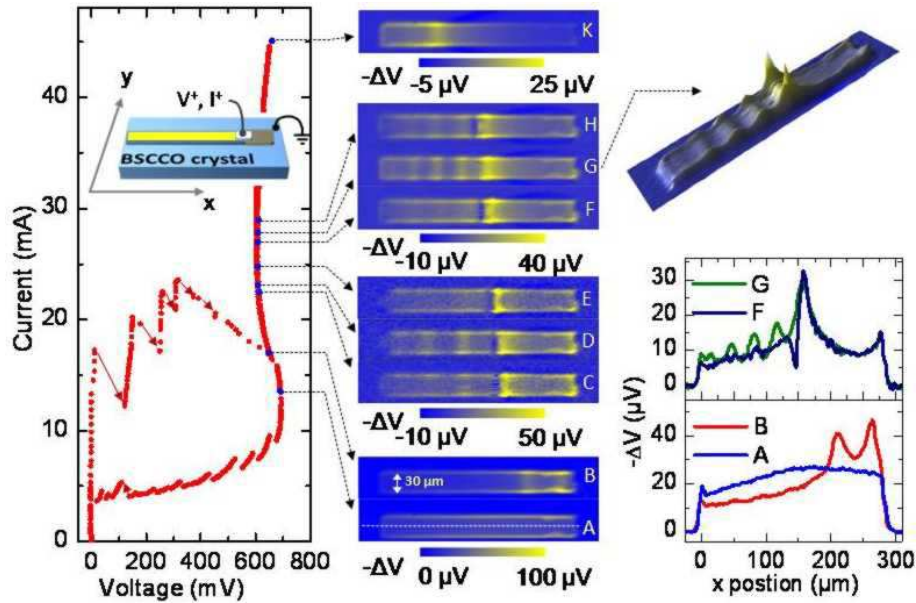}
  \caption{(double column): (color online) Current voltage characteristic and LTSLM data of a 30 x 330\,$\mu m^2$ large BSCCO mesa at 50\,K. Red solid arrows in the IVC denote switching processes, black arrows indicate bias points where LTSLM images (A) to (K) have been taken.  The upper right hand graph is a 3D view of image (G). The lower graph displays line scans along the long side of the mesa at half width, \cf white dashed line in image (A). Line scans are shown for images (A), (B), (F) and (G).}
\label{fig:3}
\end{figure*}
%
Figure \ref{fig:3} shows data for a 30 \micron\ wide mesa.
At bias currents below 16\,mA the LTSLM images, apart from edge signals, showed a broad response being strongest near the contact lead, \cf image (A) and the corresponding line scan.
At higher bias currents two bright stripes appear right of the current lead, \cf image (B).
With increasing current these two stripes move away from each other until, above a current of 22 mA, \cf image (E), the right stripe joins the edge signal while the left stripe continues to move towards the left edge of the mesa.
The feature seen here is very indicative of a hot spot.
For conventional superconducting thin films, for in-plane current flow such hot spots have been imaged by LTSEM \cite{Eichele83, Doenitz07}.
Here, when the beam touches the edge of the hot spot it increases its size, leading to a signal  $\Delta$V$>0$.
By contrast, both in the ``hot'' and in the ``cold'' regions no drastic change occurs and  $\Delta V \approx0$.
In our case both in-plane and out-of-plane currents will contribute to  $\Delta V$.
However, we still expect the signal to be strongest at the edge of the hot spot, qualitatively leading to the same pictures.
Since for a given current the c-axis resistance is lower above $T_{c}$ than below, we expect  $\Delta V <0$, as measured.
Note that the IVC shows a small jump towards lower voltages at a bias of 16 mA.
Here, the hot spot nucleates, \ie a small part of the mesa is driven to a temperature above $T_{c}$.
For mesas of larger width the hot spot feature becomes elliptic in shape, \cf Fig. S1 in Ref.\footnote{See supplementary information for additional transport and LTSLM data and an estimate of the in-phase mode velocity for large junction numbers and temperatures close to $T_{c}$}\label{footnote:1} for a 70\micron\ wide mesa.

At bias currents near 23 mA, \cf images (C) to (E), wave-like structures appear left of the hot spot boundary.
Three maxima can be seen.
The distance between them is 55\micron.
Interpreting the structures as a standing electromagnetic wave we associate a wavelength of $\lambda\approx110$\micron\ with it.
The frequency of this mode can be estimated from the total voltage drop across the mesa, $V=610$ mV, yielding    $f\approx0.44$\,THz and a mode velocity of $4.8\cdot10^{7}$\,m/s, which is about a factor of 2 lower than $c_{0}/n$.
With increasing bias current the wave feature first disappears, but, near a bias of 28 mA, is replaced by another wave-like feature exhibiting 4 maxima at a distance of 35\micron, \cf image (G).
The corresponding wave length is $\lambda\approx70$\micron\ and the calculated mode velocity is $3\cdot10^{7}$m/s.
We also found standing wave patterns for wider mesa structures, \cf Fig. S1 in Ref. [31] for a 70\micron\ wide mesa.

For an $N$ junction stack there are $N$ different mode velocities $c_{q}$
\cite{Kleiner94,Kleiner00, Kleiner01}, the fastest of which decrease inversely
proportional to the mode index $q$ counting the number of half-waves perpendicular to
the layers.
Thus, modes with, respectively, $q$ = 2 and 3 may have been excited.
However, one should also consider a mode velocity $c_{1}$ which is lower than $c_{0}/n$.
Such a scenario is realistic, since even the ``cold'' part of the mesa is likely to have a temperature not too far from $T_{c}$. In that case, the in-plane London penetration depth $ \lambda_{||}$ can become comparable to the mesa thickness $d$ and, as discussed in Ref. [31], the in-phase mode velocity becomes $c_{1}\approx0.13(c_{0}/n)(d/\lambda_{||})$, approaching zero for $T\rightarrow T_{c}$.
For $\lambda_{||}\approx 0.5$\micron\ (88 \% of $T_{c}$ for BSCCO) one obtains $c_{1}\approx3\cdot10^7$\,m/s in agreement with the mode velocity of the resonance near 28 mA.

Having interpreted the wave structure in terms of an electromagnetic wave (without having measured THz emission directly) we should also discuss alternative explanations.
In the 70\micron\ wide mesa of Fig. S1 [31] the wave structure appeared in a region of negative differential resistance.
One could assume that, like in a Gunn diode, some domains move along the junctions giving rise to the observed pattern.
However, for the case of the 30\micron\ wide mesa  shown in Fig. 1 the differential resistance was positive in the regime where the wave structures appear, making this explanation unlikely.
Further, the resonances appeared only at bath temperatures well below $T_{c}$, and a magnetic field of about 60\,G applied parallel to the layers, was sufficient to destroy the pattern, \cf Fig. S2 in Ref. [31].
All this strongly points to an interpretation in terms of the Josephson effect.
Regarding the role of the hot spot we note that its edge is typically a half wave length away from the first antinode of the wave.
The hot spot seems to have an active role in the formation of the standing wave.
As the most likely explanation its edge may be viewed as a resistive termination of the cavity formed by the cold part of the junction which is adjustable in space by applying variable values of bias current.
It thus, in a natural way, combines the approaches of shunting intrinsic junctions and using internal cavity resonances to synchronize the different junctions in the stack.
The effect may serve as an important tool to tune synchronous THz emission from intrinsic Josephson junction stacks.

Financial support by INTAS grant 05-1000008-7972 is gratefully acknowledged. HBW acknowledges the support by Special Coordination Funds for Promoting Science and Technology from the Ministry of Education, Culture, Sports, Science and Technology of the Japanese Government.

%
%
\bibliography{./Wang-hot-spots-waves-with-suppl}
%
\clearpage
\section*{Supplementary Information:}

\subsection{Transport and LTSLM data of a 70 x 330 $\mu{\rm m}^2$ large BSCCO mesa }

Figure S1 shows LTSLM and transport data for a 70\micron\ wide mesa
on a crystal with $T_{c} = 83$\,K.
For currents below 15\,mA, apart from edge signals, the only
structure is a broad spot (with  $\Delta V < 0$) near the current
lead that grows in size and intensity with increasing current, \cf
LTSLM images (A) and (B).
To understand this signal we note that the current distributes from
the contact lead across the mesa, with an in-plane current partially
carried by the contacting Au layer and partially carried by the
CuO$_2$ planes (as a result also the c-axis current density is
inhomogeneous at least in the uppermost layers where the laser beam
warms up the mesa most strongly).
When the laser beam warms up an area where the supercurrent flow
through the CuO$_2$ planes is close to critical there will be a
redistribution of currents and consequently a signal  $\Delta V \ne
0$.
By contrast, undercritical regions will not lead to a pronounced signal.

For currents above 20\,mA (where the current voltage characteristic
has a kink) an elliptic spot stronger in intensity and sharper in
profile forms in the center of the stack, \cf LTSLM image (C).
This signal is the 2D analog of the hot spot feature discussed in the
main text.
For currents above 26\,mA, a wave-like structure appears to the left
of the hot spot, consisting of two half-waves along the short side of
the mesa, corresponding to a wavelength of $\lambda_{y}\approx
70$\micron.
There is also a clear modulation along the long side; in addition a
positive signal is superimposed on the wave somewhat complicating the
determination of the wavelength (we have seen such positive signals
in several cases; the most likely explanation is that an additional
junction located close to the base crystal becomes resistive when the
laser hits its "weakest" part).
Taking the two bright regions in image (D) as the positions of the
antinodes we obtain a wavelength of $\lambda_{x}=140$\micron\ in this
direction.
>From the current voltage characteristic we find a total voltage of
670\,mV, leading to $V/N \approx 1$\,mV and a frequency of 0.5\,THz.
Using $f=c(\lambda_{x}^{-2}+\lambda_{y}^{-2})^{\frac{1}{2}}$ one
finds a mode velocity of about $3.1\cdot10^7$\,m/s, which is similar
to the mode velocities calculated for the more narrow junctions.
\begin{figure*}
  \includegraphics[width=1.8\columnwidth,clip]{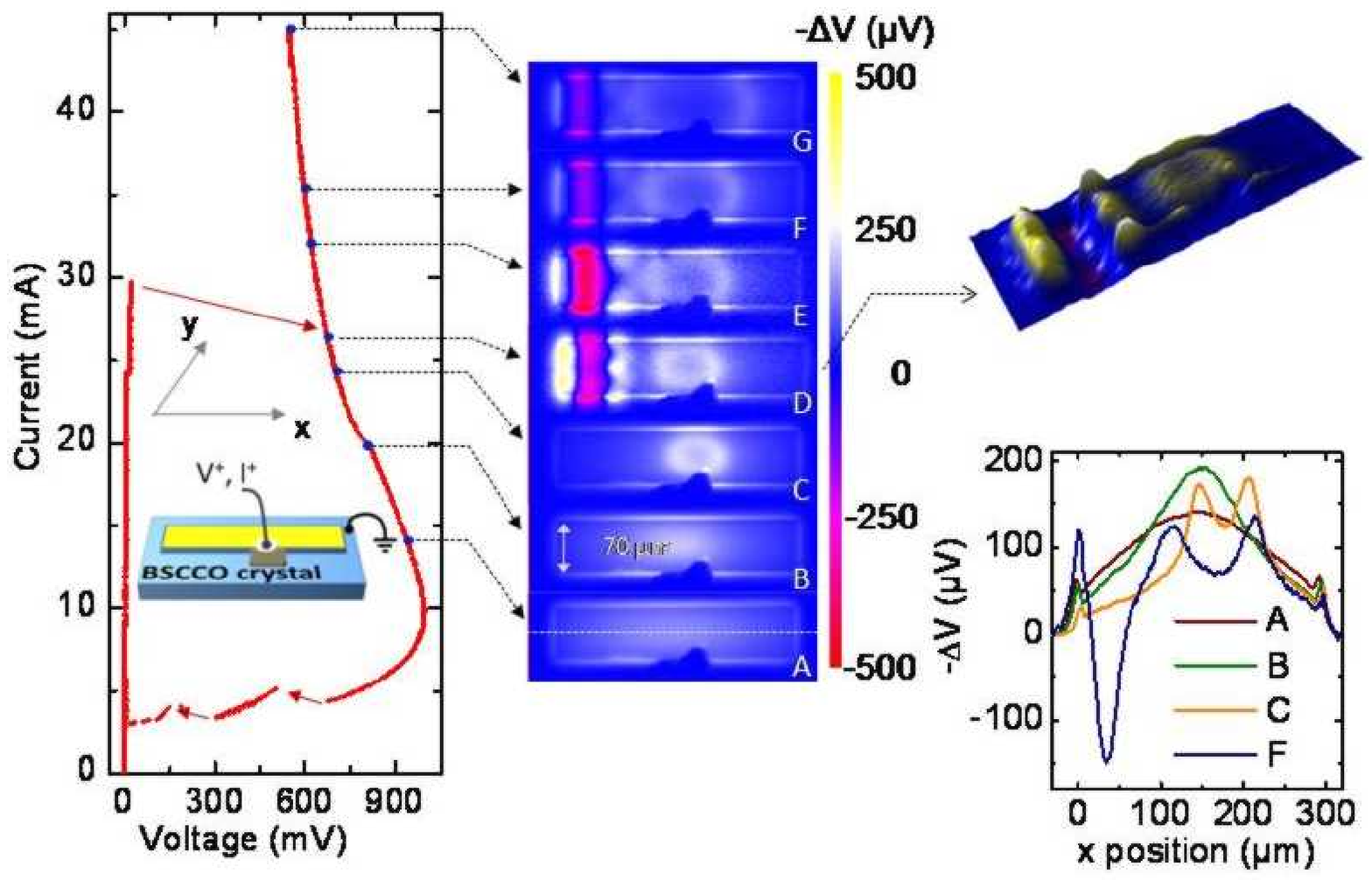}\\
{Figure S1: Current voltage characteristic (IVC) and LTSLM data of a
70 x 330 $\mu{\rm m}^2$ large BSCCO mesa at 28 K. Solid arrows in the
IVC denote switching processes, black dashed arrows indicate bias
points where LTSLM images (A) to (G) have been taken.  The upper
right hand graph is a 3D view of image (D). The lower graph displays
line scans along the long side of the mesa at half width, cf. white
dashed line in image (A). Line scans are shown for bias points (A),
(B), (C) and (F).} \label{fig:4}
\end{figure*}

\subsection{LTSLM data of a 30 x 330 $\mu{\rm m}^2$ large BSCCO mesa: effect of a magnetic field}

If the observed wave pattern is associated with the Josephson effect,
a magnetic field applied parallel to the layer should affect the
resonance.
The field should produce roughly one flux quantum per layer to significantly detune the resonance.
In Figure S2 we show that a 58\,G field, corresponding to about half
a flux quantum per junction in the ``cold'' part of the mesa is able
to completely suppress the resonance.
The data were taken for the 30 x 330 $\mu{\rm m}^2$ large BSCCO mesa
discussed in the main text.
\begin{figure*}
  \includegraphics[width=1.5\columnwidth,clip]{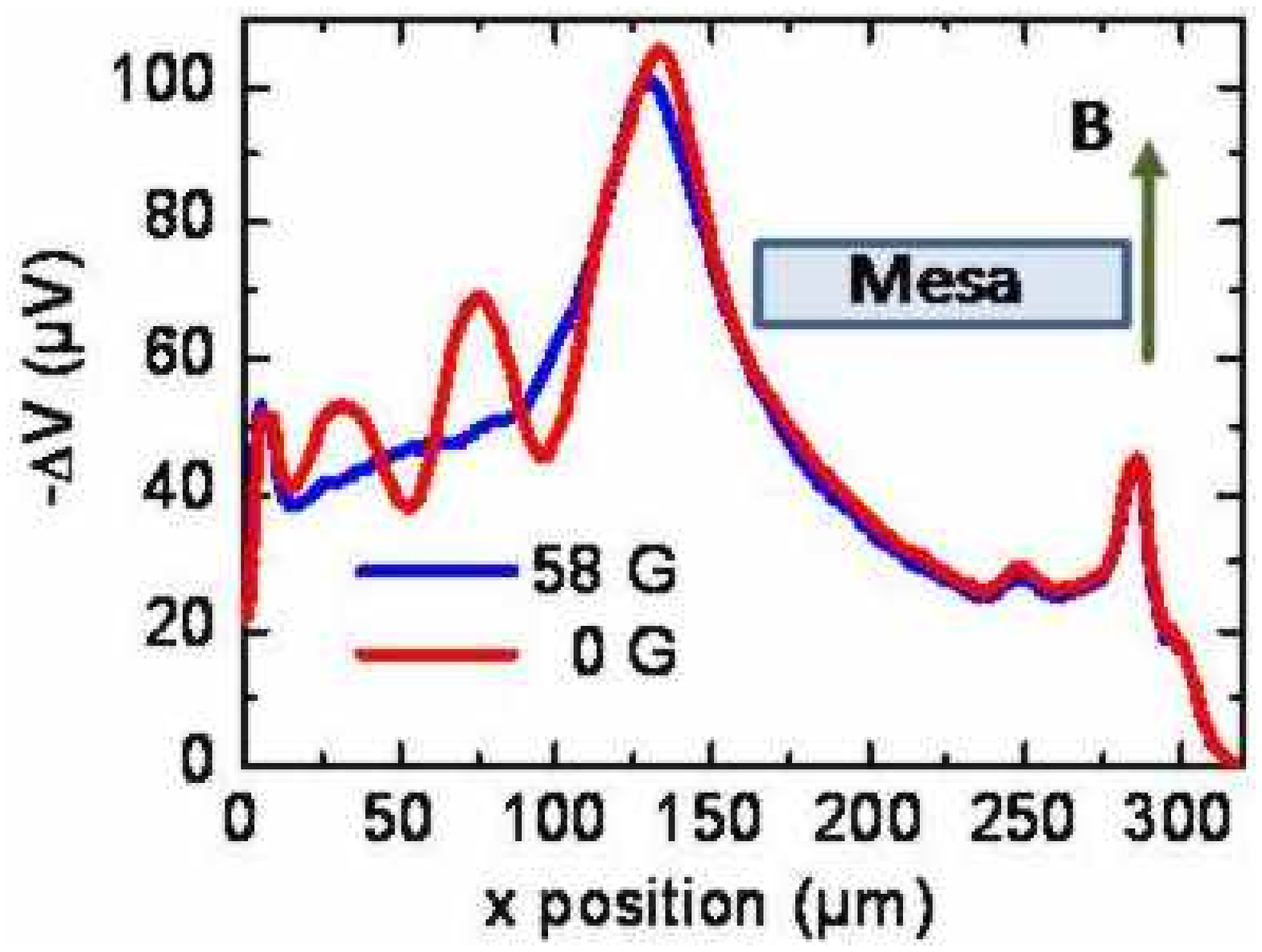}\\
Figure S2: Suppression of the standing wave pattern by a magnetic
field. The data (line scans at half width) were taken for a 30 x 330
$\mu{\rm m}^2$ large BSCCO mesa. The magnetic field was applied
parallel to the short side of the mesa (see inset). \label{fig:5}
\end{figure*}

\subsection{In-phase mode velocity for large junction numbers and temperatures close to $T_{c}$}

For an $N$ junction stack one generally finds $N$ mode velocities [1,
2] $c_{q}=\omega_{pl}\lambda_{J}/\sqrt{1-2\tilde s \cos\{\pi
q/(N+1)\}}$ , with $q = 1...N$.
Here, $\omega _{pl}$ denotes the Josephson plasma frequency and
$\lambda_{J}$ is the Josephson penetration depth; $\tilde s$ is the
inductive coupling parameter and can be expressed via $1-2\tilde{s}=
(\lambda_{j}/\lambda_{\bot})^2$, where $\lambda_{\bot}$ is the
out-of-plane penetration depth.
When the thickness $d'$ of the superconducting layers is much smaller
than the in-plane London penetration depth $\lambda_{||}$, for the
product $\bar{c}=\omega_{pl}\lambda_{J}$ one finds
$\bar{c}=\omega_{pl}\lambda_{J} =
(c_{0}/n)\sqrt{td'/2}/\lambda_{||}$, where $t$ is the barrier
thickness and $n$ is the refractive index.
We will further make use of the relation
$\lambda_{J}=\frac{\lambda_{\bot}}{\lambda_{||}}\cdot\sqrt{sd'/2}$,
where s (=1.5 nm for BSCCO) is the layer period.

For $N >> 1$ and $(\lambda_{J}/\lambda_{\bot})^2 << 1$ one finds for
the in-phase mode velocity $c_{1}$:
\begin{eqnarray*}
c_{1} \approx \frac{\omega_{pl}\lambda_{J}}{\sqrt{(1-2\tilde s)+\tilde s \pi^2/N^2}}=\frac{\omega_{pl}\lambda_{J}}{\sqrt{(\lambda_{J}/\lambda_{\bot})^2+\tilde s \pi^2/N^2}}\\ =\frac{\omega_{pl}\lambda_{J}}{\sqrt{(sd'/2\lambda_{||}^2)+ \pi^2/2 N^2}}
\end{eqnarray*}
The first term under the square root dominates for
$d/\lambda_{||}>\pi\sqrt{s/d'}$, where $d =Ns$ is the thickness of
the mesa.
In this case one finds $c_{1} \approx (c_{0}/n)\sqrt{t/s}\approx
(c_{0}/n)$.
For BSCCO, $d'\approx 0.3$\,nm is used often, yielding
$\pi\sqrt{s/d'}\approx 7$.
Note that, since $d'$ is not defined very well, this number should
not be taken too literally; if $d'$ is closer to the interlayer
distance $s$ the ratio $d/\lambda_{||}$ is closer to $\pi$.
Using $\lambda_{||}(T=0)\approx 170$\,nm, $d$ must be of order
1\micron\ to fulfill that.
If the second term under the square root dominates, one obtains
$c_{1} \approx \sqrt{2}N\bar{c}/\pi =
(c_{0}/n)d\sqrt{td'}/s\pi\lambda_{||}\approx
0.15(c_{0}/n)(d/\lambda_{||})$.
Since $\lambda_{||}$ diverges for $T \rightarrow T_{c}$ we see that
$c_{1}$ approaches zero for $T \rightarrow T_{c}$.
For $\lambda \approx 0.5$\micron\ (88\% of $T_{c}$ for BSCCO,
assuming a Ginzburg-Landau temperature dependence for the in-plane
penetration depth, $\lambda_{||} \propto 1/\sqrt{1-T/T_{c}}$) one
obtains $c_{1}\approx 0.3c_{0}/n$, which is comparable to the mode
velocities estimated in the main text.

\subsection{References}

$[1]$ R. Kleiner, Phys. Rev. B \textbf{50}, 6919 (1994); R. Kleiner, T. Gaber, and G. Hechtfischer, Phys. Rev. B \textbf{62}, 4086 (2000).\\
$[2]$ S. Sakai, A.V. Ustinov, H. Kohlstedt, A. Petraglia, and N.F. Pedersen, Phys. Rev. B \textbf{50}, 12905 (1994).
\end{document}